\documentclass[preprint]{iucr}              

     \journalcode{A}              
\usepackage{amsmath,amsfonts,amssymb} 
 
\begin{document}

\title{Time-dependent dynamical Bragg diffraction in deformed crystals by beam propagation method (BPM)}

\cauthor[a]{Jacek}{Krzywinski}{krzywins@slac.stanford.edu}{}
\author[a]{Aliaksei}{Halavanau}

\aff[a]{SLAC National Accelerator Laboratory, Stanford University, Menlo Park CA 94025, USA}

\maketitle

\begin{synopsis}
We present a novel method of simulating dynamic diffraction in distorted crystals using fast Fourier transform-based beam propagation approach.
\end{synopsis}

\begin{abstract}
This paper describes how to efficiently solve time-dependent x-ray dynamic diffraction problems in distorted crystals with a fast Fourier transform-based beam propagation method (FFT BPM). We show examples of using the technique to simulate the propagation of x-ray beams in deformed crystals in space and time domains relevant to the cavity-based x-ray Free Electron Lasers (CBXFELs) and XFEL self-seeding systems. 
\end{abstract}


\section{INTRODUCTION}
\label{sec:intro}  

The Beam Propagation Method (BPM) is a popular method to simulate the propagation and scattering of waves in a non-homogeneous media  \cite{Ersoy2007} and has been frequently used to design and simulate optoelelctronic devices \cite{Sziklas75,Feit78,SHAABAN2019,Hadley92,Yamauchi1995,SCALORA1994191}. The BPMs can be roughly divided into two groups. The first group employs finite difference numerical schemes and is called FD BPM. The FD BPM methods are mainly used for problems where the radiation wavelength is comparable with one of the dimensions of the simulated elements or with the length of the pulse. The second group, FFT BPM, uses fast Fourier transforms (FFT) to solve the governing wave equations. The FFT BPM is a method of choice when the slowly varying envelope approximation (SVEA) can be applied. The popularity of the FFT BPM is attained by its relatively simple algorithm. The advantage of the FFT BPM is that it is formulated as an initial value problem. Therefore, the boundary conditions are automatically fulfilled. It is straightforward to model complicated optical elements by implementing a spatial distribution of dielectric susceptibility. The FFT BPM method is also well suited for modeling problems with highly coherent x-ray beams produced by x-ray free-electron lasers (XFELs) or 4-th generation synchrotron radiation sources. The small value of x-ray susceptibility leads to further simplifications of the governing equations, especially when para-axial approximation is justified. There is an increasing interest in applying FFT BPM for simulating  propagation of highly coherent x-ray beams through complicated x-ray optics \cite{Gaudin12,ANDREJCZUK2015,Morgan2015,bajtAIP,Li2017,articleYan,pfeiffer,PhysRevLett.125.254801, Halavanau15511}.
This paper will focus on FFT BPM, which could be efficiently applied in time-dependent simulations of diffracted XFEL pulses from deformed crystals of arbitrary shapes.

\section{Theoretical approach}
\label{sec:Theo}  
The scalar Helmholtz equation for a constant angular  frequency $\omega$:
\begin{equation}
\label{eq:fov1}
		\left(\nabla^2+k(n(x,y,z))^2\right)E(x,y,z)=0
\end{equation}
 can be written in the following form\cite{Hadley92,Ersoy2007}:
 \begin{equation}
\label{eq:fov2}
	\frac{d\psi}{d\hat{z}}=i\left(\sqrt{1+{\nabla_\bot}^2-\delta\epsilon(\hat{x},\hat{y},\hat{z})}-1\right)\psi
\end{equation}
 where $\hat{x},\hat{y},\hat{z}= kx,ky,kz $, $k$ is the wave vector,${\nabla_\bot}^2$ is the Laplace operator taken with respect to the transverse coordinates $\hat{x}, \hat{y}$, $  \delta\epsilon(\hat{x},\hat{y},\hat{z})=n(\hat{x},\hat{y},\hat{z})^2-\bar{n}^2 $, $n(\hat{x},\hat{y},\hat{z})$ is the refractive index, $\bar{n}$ is its average value and  $ E(x,y,z)=\psi(x,y,z)e^{-ikz}$.
In the case of  beams with narrow angular spectrum this equation can be approximated by:
 \begin{equation}
\label{eq:fov3}
	\frac{d\psi}{d\hat{z}}=i\left(\sqrt{1+{\nabla_\bot}^2}-\frac{\delta\epsilon(\hat{x},\hat{y},\hat{z})}{2\sqrt{1-\left\langle k_x^2+k_y^2 \right\rangle}}-1\right)\psi
\end{equation}	

Here the operator $\sqrt{1+{\nabla_\bot}^2}$ in the denominator of the second term of the r.h.s  was approximated by $\sqrt{1-\left\langle k_x^2+k_y^2 \right\rangle}$ where $\left\langle k_x^2+k_y^2 \right\rangle$ 
is the  average angular spectrum and $k_x^2+k_y^2=F[{\nabla_\bot}^2]$ is the Fourier transform of  the ${\nabla_\bot}^2$ operator.
Eq. \eqref{eq:fov3} leads to the following FFT BPM equation:
 \begin{equation}
\label{eq:fov4}
	\psi(\hat{x},\hat{y},\hat{z}+\Delta{\hat{z}}) \approx F^{-1}\left\{F\left[\psi(\hat{x},\hat{y},\hat{z})\right]e^{i\Delta{\hat{z}}\sqrt{1-k_x^2-k_y^2}}\right\}e^{i\Delta{\hat{z}}\frac{\delta\epsilon(\hat{x},\hat{y},\hat{z})}{2 \cos{\alpha}}}
\end{equation}
where $F$ and $F^{-1}$ denote Fourier and inverse Fourier transforms, $\cos{\alpha}=\sqrt{1-\left\langle k_x^2+k_y^2 \right\rangle}$. The angle $\alpha$ corresponds to the grazing incidence angle with respect to the propagation direction.  
We have previously applied Eq. \eqref{eq:fov4} to take into account dynamic diffraction effects when simulating  interaction of x-ray beams with gratings and multilayer optics \cite{Gaudin12,ANDREJCZUK2015,Morgan2015,bajtAIP}. 

   \subsection{Application to crystal diffraction}
   
   In a large class of dynamical diffraction problems, the angular spectrum of the scattered x-ray beams consists of two narrow bands centered around the incident and the reflection angles. The meaningful information is contained within these bands. One can remove the fast oscillating component related to inter-planar spacing from the physical picture and derive FFT BPM equations for slowly varying envelopes of transmitted and reflected beams. For simplicity, we are considering the 2D ($x,z$) case, and the generalization to the 3D case is straightforward. The procedure is outlined below. First, one can write the scattered x-ray field as a sum of two components:   
  \begin{equation}
\label{eq:fov6}
    	\psi(\hat{z},\hat{x})=\psi(\hat{z},\hat{x})_{+} + \psi(\hat{z},\hat{x})_{-}.
\end{equation}
  Then the slowly varying envelopes are defined as follows:
  \begin{equation}
\label{eq:fov7}
    	\widetilde{\psi}(\hat{z},\hat{x})_{+} = \psi(\hat{z},\hat{x})_{+}e^{-i\frac{k_{d}}{2}\hat{x}},\ \  \widetilde{\psi}(\hat{z},\hat{x})_{-} = \psi(\hat{z},\hat{x})_{-}e^{i\frac{k_{d}}{2}\hat{x}}
\end{equation}
where $k_d$ is the reciprocal vector related to inter-planar spacing.
Next we expand the exponential term related to dielectric susceptibility in Eq. \eqref{eq:fov4} as:
  \begin{equation}
\label{eq:fov9}
    e^{i\hat{z}\frac{\delta\epsilon(\hat{z},\hat{x})}{2cos{\alpha}}} = \sum_{n=-\infty}^{+\infty}\Delta\epsilon(\hat{z},\hat{x})_{n}e^{ik_{d}n\hat{x}}
\end{equation}
When crystal planes, taking part in the reflection, are perpendicular to the $x$-axis, one can describe the $\delta\epsilon(\hat{z},\hat{x})$ in Eq. \eqref{eq:fov9} as a complex harmonic function with the period equal to the inter-planar  spacing:
  \begin{equation}
\label{eq:fov8}
    \delta\epsilon(\hat{z},\hat{x}) = \chi_{0}+2\chi_h\cos{k_d\hat{x}}
\end{equation}
and $\chi_{0}=\chi_{0r}+i\chi_{0h}$ and $\chi_h=\lvert\chi_{rh}\rvert-i\lvert\chi_{ih}\rvert$.
The complex values of $\chi_{0r}$, $\chi_{0h}$, $\chi_{rh}$ and $\chi_{ih}$ could be found, for example, at the web page “x-ray dynamical diffraction data on the web” (https://x-server.gmca.aps.anl.gov/x0h.html) or XOP x-ray optics software toolkit \cite{XOPcode}.

By applying the Jacobi–Anger expansion: $e^{i w \cos{\theta}} = \sum_{n=-\infty}^{\infty} i^n J_n(w) e^{i n \theta}$  the first three expansion terms in Eq. \eqref{eq:fov9} can be written in the following form: 
  \begin{equation}
\label{eq:fov11}
	\Delta\epsilon_{0}=J_{0}\left( \hat{z}\frac{\chi_h}{\cos{\alpha}}\Pi(\hat{x},\hat{z})\right)e^{\frac{i\hat{z}}{2\cos{\alpha}}\chi_{0}\Pi(\hat{x},\hat{z})}
\end{equation}

  \begin{equation}
\label{eq:fov12}
	\Delta\epsilon_{1}=\Delta\epsilon_{-1}=iJ_{1}\left(\hat{z}\frac{\chi_h}{\cos{\alpha}}\Pi(\hat{x},\hat{z})\right)e^{\frac{i\hat{z}}{2\cos{\alpha}}\chi_{0}\Pi(\hat{x},\hat{z})}
\end{equation}
where $\Pi(\hat{x},\hat{z}) = 1$ inside the crystal,  and $\Pi(\hat{x},\hat{z}) = 0$ outside the crystal, with $J_n(w)$ being Bessel functions. 

 We treat deformation of the crystal as modification of the susceptibility \cite{Authier2003-ae}:
 
   \begin{equation}
\label{eq:Authier}
    	\Delta\epsilon^{'}_{n}(\hat{z},\hat{x}) = \Delta\epsilon_{n}(\hat{z},\hat{x})e^{in\mathbf{k_{d}u}(\hat{z},\hat{x})}
\end{equation}
 where $\mathbf{k_{d}u}$ is a scalar product of the reciprocal  and displacement vectors. We note that  $\mathbf{k_{d}}$ is parallel to $\hat{x}$ by definition, therefore only $x$-component of $\mathbf{k_{d}}$ is used. 
Inserting Eq. \eqref{eq:Authier} into Eq. \eqref{eq:fov4} and neglecting terms proportional to $e^{ik_{d}n\hat{x}}$ for $\left|n\right|>1$ one arrives at the following set of equations for slowly varying envelopes $\widetilde{\psi}(\hat{z},\hat{x})_{+}$ and $\widetilde{\psi}(\hat{z},\hat{x})_{-}$: 

  \begin{equation}
  \begin{aligned}
\label{eq:fov10}
    \widetilde{\psi}(\hat{z}+\Delta\hat{z},\hat{x})_{+} =  \widetilde{\psi}(\hat{z},\hat{x})_{p+}\Delta\epsilon_0(\hat{z},\hat{x})+\widetilde{\psi}(\hat{z},\hat{x})_{p-}\Delta\epsilon^{'}_{+1}(\hat{z},\hat{x}) \\ 
        \widetilde{\psi}(\hat{z}+\Delta\hat{z},\hat{x})_{-} =  \widetilde{\psi}(\hat{z},\hat{x})_{p-}\Delta\epsilon_0(\hat{z},\hat{x})+\widetilde{\psi}(\hat{z},\hat{x})_{p+}\Delta\epsilon^{'}_{-1}(\hat{z},\hat{x}),
    \end{aligned}
\end{equation}
where $\widetilde{\psi}(\hat{z},\hat{x})_{p\pm}=F^{-1}\left[F\left[\widetilde{\psi}(\hat{z},\hat{x})_{\pm}\right]p_{\mp}\right]$.
The operators $p_{\pm}=e^{i\Delta{\hat{z}}\sqrt{1-(k_{x} \pm \frac{k_d}{2})^2}}$ correspond to the Fourier image of ${\nabla_\bot}^2$ which is shifted by $\pm \frac{k_d}{2}$ in the angular spectrum space.
 Equation \eqref{eq:fov4} can be rewritten in the operator form as: 
\begin{eqnarray}
\label{split-op}
    \psi(\hat{x},\hat{y},\hat{z}+\Delta{\hat{z}}) &\approx& \prod_i A(a_i \Delta_z) B (b_i \Delta_z) \psi(\hat{x},\hat{y},\hat{z}), \\ \nonumber
    A &=& F^{-1}\left\{F\left[\psi(\hat{x},\hat{y},\hat{z})\right]e^{i a_i\Delta{\hat{z}}\sqrt{1-k_x^2-k_y^2}}\right\},\\ \nonumber
    B &=& e^{i b_i\Delta{\hat{z}}\frac{\delta\epsilon(\hat{x},\hat{y},\hat{z})}{2 \cos{\alpha}}}.
\end{eqnarray}
In the case of first order splitting, when $a_1 = 1, b_1 = 1$, Eq. \eqref{split-op} is reduced to Eq. \eqref{eq:fov4}. For most of the practical problems this splitting scheme is accurate. Higher order splitting improves the efficiency of the FFT BPM for large Bragg angles. For instance, 
when $a_1 = 0.0, a_2 = 1.0, b_1 = 0.5, b_2 = 0.5$, Eq. \eqref{split-op} is accurate up to $O(\Delta z^2)$, and when $t = 1.3512, a_1 = 0.0, a_2 = t, a_3 = 1 - 2t, a_4 = t, b_1 = t/2, b_2 = (1 - t)/2, b_3 = (1 - t)/2, b_4 = t/2$ it is accurate up to $O(\Delta z^4)$; see \cite{OMELYAN2002188}.

We found numerically that we can considerably improve the efficiency in the split-operator scheme if we introduce a Zassenhaus-like exponent \cite{CASAS20122386} in Eq. \eqref{eq:fov10}:
\begin{equation}
   \Delta \epsilon_0 \rightarrow \Delta \epsilon_0 e^{\frac{D_2}{2}\Delta z^2}, \quad  D_2=(\chi_h / 2.0 / \cos{\alpha})^2
\end{equation} 
This exponent allows for much larger $\Delta z$ steps and faster computation while maintaining numerical accuracy. We note that it is a common numerical method in computational quantum mechanics.

We point out that Eqs. \eqref{eq:fov10} can be treated as a two-beam approximation for the FFT BPM. These equations are analogous to the Takagi-Taupin equations (TTE) in two-beam approximation \cite{Authier2003-ae}. However, there is a significant difference between TTE and FFT BPM equations. The TTE equations are a system of hyperbolic equations where the second derivatives in the transverse direction with respect to beam propagation are neglected. Therefore, diffraction of the x-rays is not taken into account in the TTE formulation. The numerical algorithms for solving TTE, which are presented in literature, typically require setting the boundary conditions. This could be a difficult problem itself in complicated geometries or when the boundaries are not very well defined. On the other hand, the FFT BPM equations are a system of parabolic equations that automatically includes diffraction. Also, as we have mentioned before, the FFT BPM method is especially convenient when dealing with complicated shapes or non-homogeneous boundaries. 

 The Eqs. \eqref{eq:fov10} are derived under the assumption that the crystallographic planes, taking part in the reflection, are always perpendicular to the $x$-axis. Different reflection geometries, such as symmetric or asymmetric Bragg or Laue reflections, can be easily implemented in FFT BPM by setting appropriate boundaries of the crystal's surfaces. Let us consider a crystal sample in the form of a slab. The symmetric Bragg reflection will be described by Eqs. \eqref{eq:fov10} if the $s$-axis, normal to the crystal's surfaces, is parallel to the $x$-axis. The asymmetric reflection geometries correspond to the situation when the $s$-axis is tilted with respect to the $x$-axis. 
 
 Our method does not work for the exact back-scattering geometry, as it will result in large oscillations of the exponential term in Eq. \eqref{eq:fov7}. However, we will show in the next paragraph that we are still able to obtain very accurate results for Bragg angles as large as 79 degrees. 
In the following part of the article we will focus on the symmetric Bragg reflection cases.

 \section{Numerical examples}

  \subsection{Stationary problems}
 
  \begin{figure} 
   \includegraphics[height=6cm]{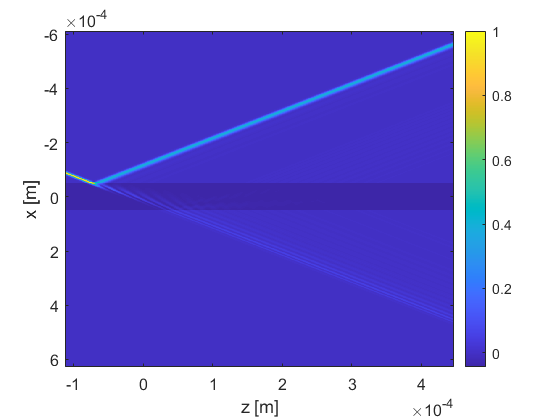}
    \includegraphics[height=6cm]{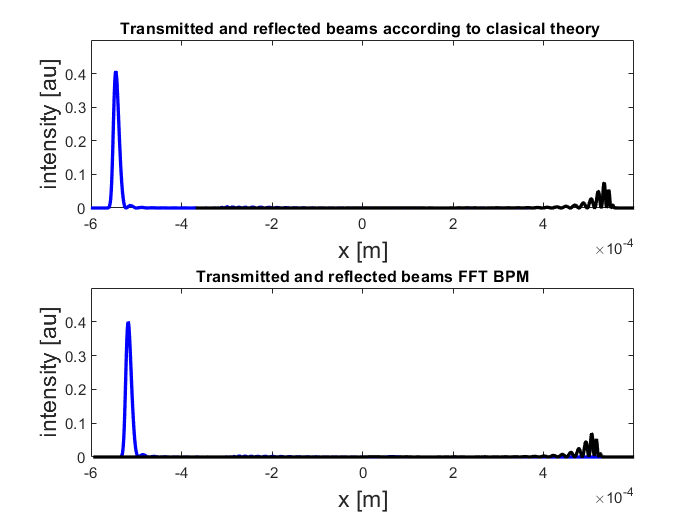}

   \caption{Top: 2D visualization of a $C^*(400)$ at 9.831 keV Bragg reflection. The incident beam has a transverse RMS (root mean square) size of 3.5 $\mu$m which is smaller than the extinction length. Bottom: comparison of the of transmitted and reflected transverse intensity profiles obtained from FFT BPM and classical diffraction theory \cite{PhysRevSTABYuri}.  }
   \label{fig:comparison}
   \end{figure}
   
   \begin{figure} 
 \centering
   \includegraphics[width=0.7\linewidth]{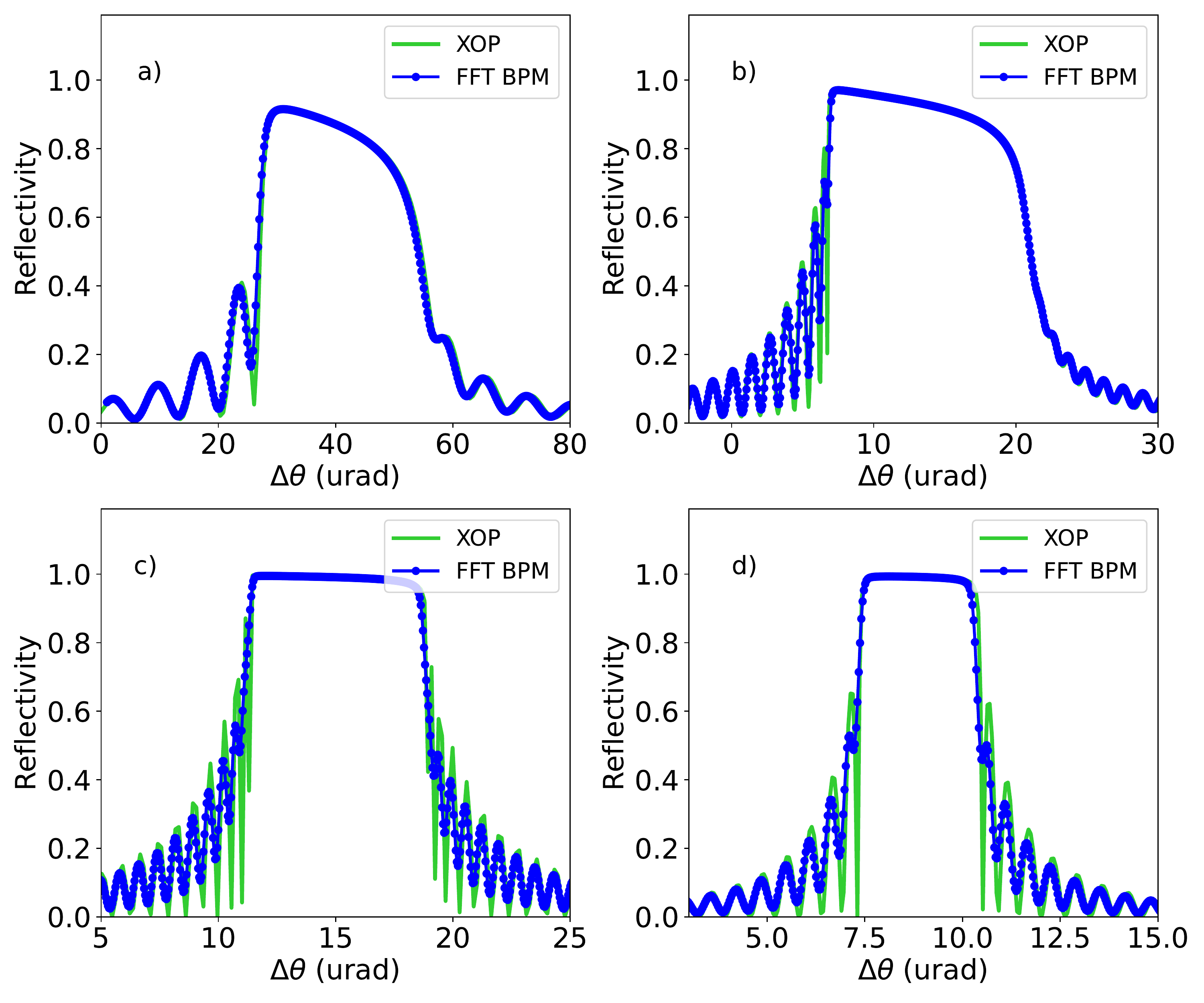}

   \caption{
a) Si (444) reflection at 8.048 keV, $\theta_B=79^\circ$, $d=50\, \mu$m b) Si (400) reflection at 9 keV, $\theta_B=30.5^\circ$, $d=50\, \mu$m c) $C^*$(400) reflection at 9.831 keV,  $\theta_B=45^\circ$, $d=100 \, \mu$m d) $C^*$(333) reflection at 12.8 keV,  $\theta_B=44.9^\circ$, $d=100\,\mu$m.
} 
\label{fig:rc}
   \end{figure}
   
 We first applied our implementation of  FFT BPM based on expansions from Eqs. \eqref{eq:fov11}, \eqref{eq:fov12} to simulate $C^*(400)$ reflection and transmission of 9.831 keV x-ray beam scattered from a 100 $\mu$m thick crystal. We have chosen the RMS transverse size of the Gaussian x-ray beam to be 3.5 $\mu$m, smaller than the extinction length, in order to observe both the reflected and the transmitted beams. We have compared results of our FFT BPM simulation with classical dynamic diffraction theory \cite{PhysRevSTABYuri} and found a very good agreement; see Fig.\ref{fig:comparison}.
 We have also checked the accuracy of our method by comparing rocking curves obtained by the FFT BPM method and the XOP software toolkit results. 
 For this test, we consider a planar radiation intensity distribution:
 \begin{equation}
     \psi^i_{xy}(x, y) = \frac{1}{(1 + i z_x/z_R)} e^{-\frac{(x - x_0)^2 + y^2}{W_0^2(1 + i z_x/z_R)}} e^{i x \sin{\theta_i}}
 \end{equation}
incident on a crystal, where $z_x$ is distance to the source point, $z_R$ is the Rayleigh range, $W_0$ is the Gaussian waist size and $\theta_i$ is the incidence angle.
 The example results simulated for Silicon and Diamond crystals are illustrated in Fig. \ref{fig:rc}. We point out a good agreement with XOP. This holds true even when the Bragg angle approaches the back-scattering geometry, with the Bragg angle $\theta_B=79^\circ$, as it is shown in  Fig. \ref{fig:rc} a). We have noticed that for the the Bragg angle $\theta_B=79^\circ$ the second order splitting scheme worked more efficiently than the first order splitting scheme. The first order splitting scheme is more efficient for the other cases shown in Fig. \ref{fig:rc}. The slight remaining discrepancies between FFT BPM and XOP are attributed to the fact that in our simulations we use finite Gaussian beams with angular divergence while XOP employs plane waves. The latter fact illustrates another advantage of FFT BPM in cavity-based systems with thin crystals, where the accuracy of modelling the spectral/temporal field profile is crucial.

  \subsection{Deformed crystals}
\begin{figure}
   \begin{center}
   \includegraphics[width=0.46\linewidth]{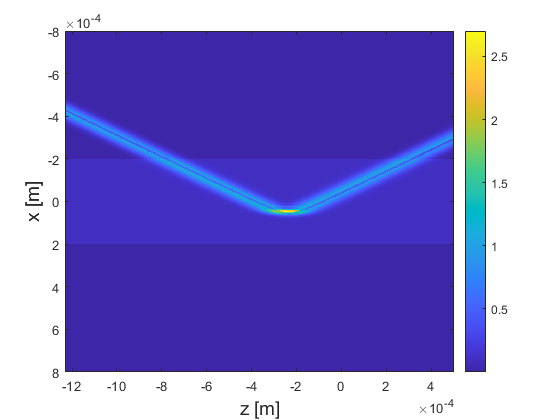}
      \includegraphics[width=0.46\linewidth]{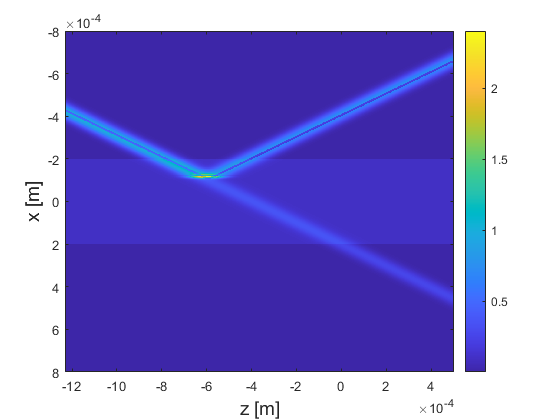}

   \end{center}
 
   \caption{2D visualization of a Si(444) Bragg reflection with a constant strain gradient in the crystal for two different strain gradient parameters: B= 0.05 $m^{-1}$ (left) and 0.25 $m^{-1}$ (right). The trajectories predicted by the geometrical PP theory are marked with dark lines. Color scale units are arbitrary.}
   \label{fig:stress}
   \end{figure}

 Let us now consider crystal deformation in the form of a constant strain gradient. This is one of the few instances when a dynamical diffraction problem, involving deformed crystals, can be studied analytically \cite{Balibar:a21954}. We use this case to benchmark our method when applied to deformed crystals. The deformation can be easily described by the distribution of $\Delta\epsilon_n^{'}(\hat{z},\hat{x})$, n=-1,0,1 given by Eq. \eqref{eq:Authier}. For the constant strain gradient case it has the following form:
    \begin{equation}
\label{eq:Authier2}
    	\Delta\epsilon^{'}_{n}(\hat{z},\hat{x}) = \Delta\epsilon_{n}(\hat{z},\hat{x})e^{{k_{d}B\hat{x}^2}},
\end{equation}
where $B$ ($1/m$) is a constant defining the strain gradient in inverse meters.
 The influence of the constant strain gradient on the reflection can be investigated using a geometric Penning and Polder (PP) theory \cite{Polder:a04290, Gronkowski, Yan:fe5010}. The hyperbolic trajectory of the x-ray beam scattered from the strained crystal, calculated analytically from the PP theory, is marked by the solid line on the left side of Fig. \ref{fig:stress}. Here we have chosen a Si(444) reflection at 17 keV photon energy to be compatible with the study presented in \cite{Gronkowski}. The trajectory derived from the PP theory, for the parameter B equal to 0.05 $m^{-1}$, follows very closely the intensity distribution obtained from the FFT BPM numerical simulation.

The PP theory works well only if the strain gradient is much smaller than a certain critical value \cite{Gronkowski}. The effect of the strain on x-ray wavefields can be characterized using so called the strain gradient parameter $\beta$ \cite{Gronkowski}. For the symmetric Bragg case the the  strain gradient parameter $\beta$ is equal to \cite{Yan:fe5010}:
 \begin{equation}
\label{eq:Yan}
    	\beta = 4B \sin^{2}{\theta_B} \tan{\theta_B}/|\chi_h|
\end{equation}

The critical strain gradient parameter is defined as in \cite{Gronkowski} $\beta_c = \pi/(2\Lambda_0)$ where $\Lambda_0=\lambda \cos{(\theta_B)}/|\chi_h|$ is the extinction length. When the parameter $\beta$ in Eq. \eqref{eq:Yan} approaches the critical value $\beta_c$ the geometrical PP theory breaks down because new wavefields are created in the process called the interbranch scattering \cite{Balibar:a21954}. This phenomenon is illustrated on the right side in Fig. \ref{fig:stress}. Here, the parameter B is equal to 0.25 and the ratio $\beta/\beta_c=4.7326$. One can observe a newly created branch of x-ray wavefields  propagating along the incidence angle. The analytical theory \cite{Balibar:a21954} predicts that the ratio of the intensity of this new branch with respect to the incident intensity is equal to $I/I_0=\exp{(-2\pi\beta_c/|\beta|)}$ when absorption in the crystal is neglected. The ratio $I/I_0$ calculated from the FFT BPM numerical simulation agrees with the theoretical prediction at the accuracy level better than 0.1 $\%$. That gives us confidence that the FFT BPM approach can be used to model dynamical diffraction phenomena in highly strained crystals. 
    \begin{figure}
   \begin{center}
  
      \includegraphics[width=0.46\linewidth]{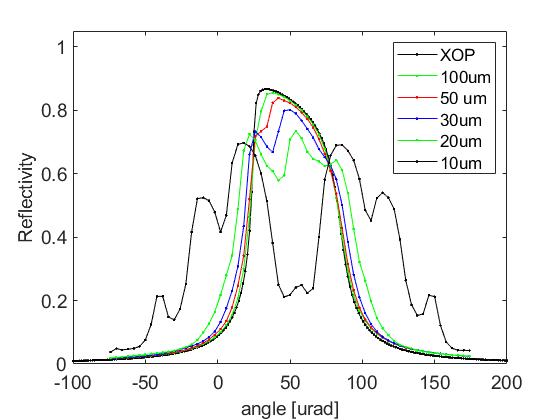}

   \end{center}
 
   \caption{ Rocking curves simulated for a crystal with sinusoidal deformation having a constant amplitude of 1 Angstrom and different periods. }
   \label{fig:sin_deform}
   \end{figure}
   
Another example involving different type of deformation is presented in Fig \ref{fig:sin_deform}. Here it has a form of sinusoidal variations along x-axis of 1 Angstrom amplitude, with different periods. We have simulated the rocking curves as a function of undulation period for the Si (111) reflection at 5 keV photon energy. The crystal was illuminated by a Gaussian beam with a waist of $W_0$=85.4 $\mu$m and located $z_x$ = 181.1 m away.

\subsection{Time-dependent problems: Ideal crystals }
\begin{figure}
   \begin{center}
   \includegraphics[width=0.69\linewidth]{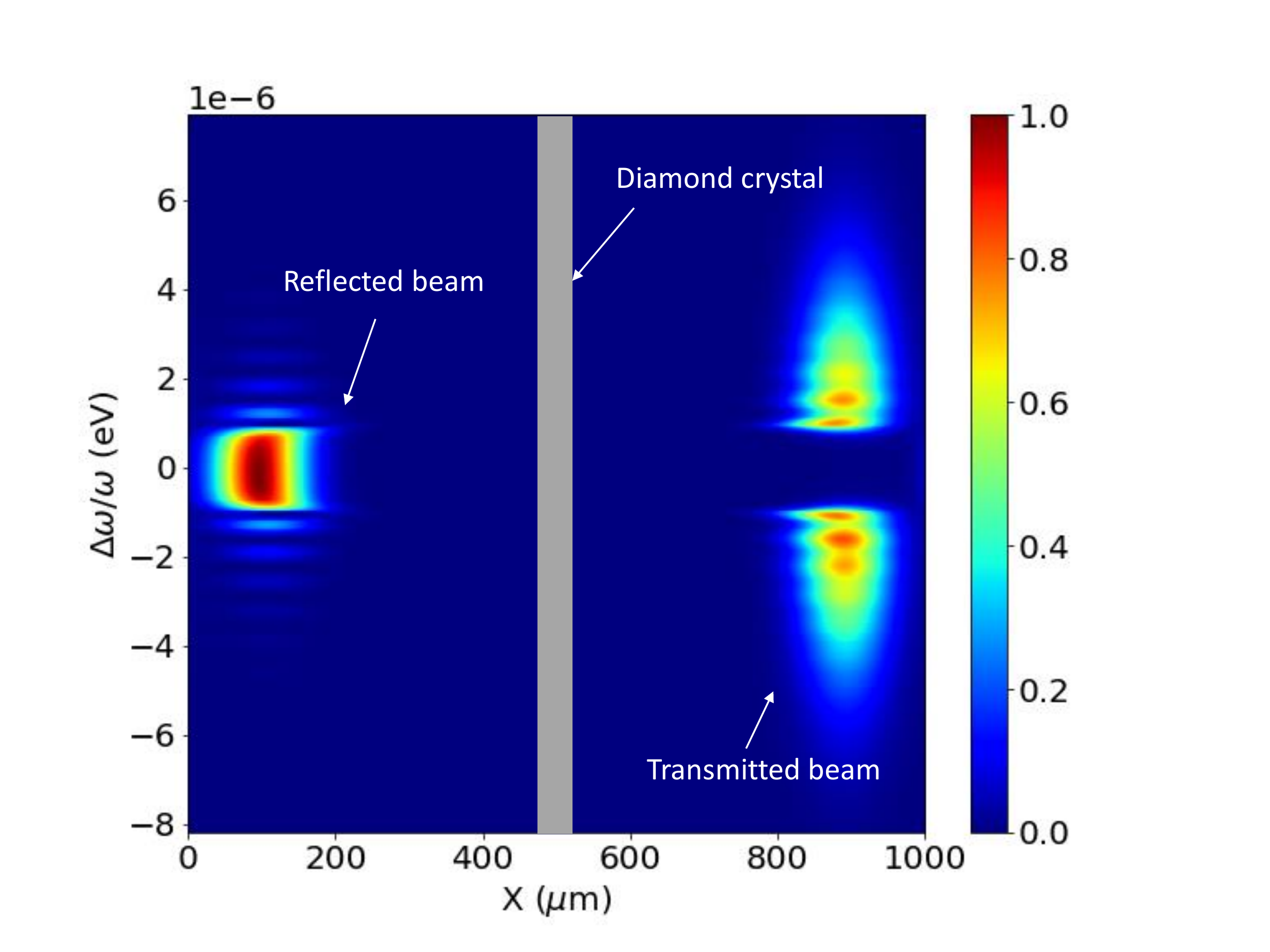}
   \end{center}
  \caption{ Spatio-spectral intensity profiles of reflected and transmitted beams for 50 $\mu$m thick crystal Diamond $C^*(333)$ at $y=0$. The incident Gaussian beam has the following  parameters: $\sigma_t = 10$ fs, $W_0 = 250$ $ \mu$m, $\omega_0=12.8$ keV. Color scale units are arbitrary.  }
  \label{fig:E-wx}
   \end{figure}

\begin{figure}
   \begin{center}
   \includegraphics[width=0.46\linewidth]{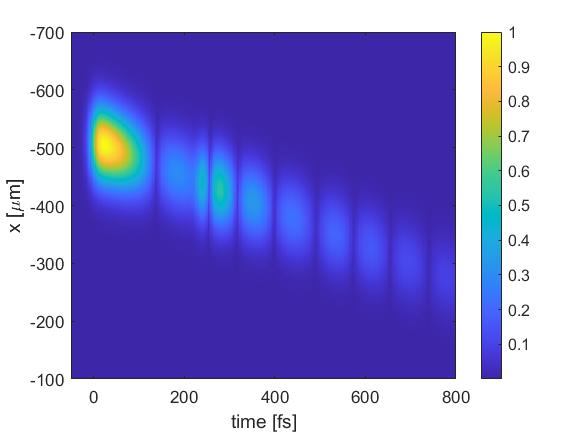}
      \includegraphics[width=0.45\linewidth]{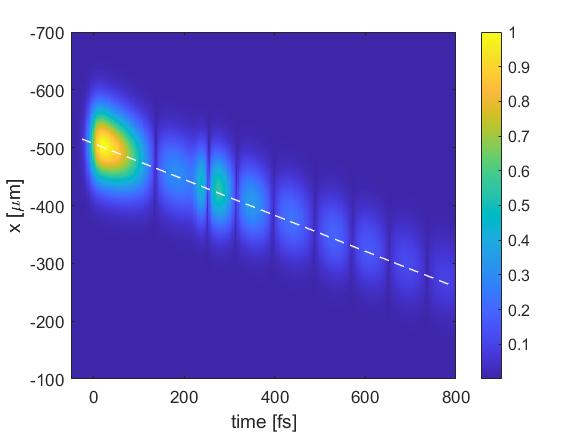}
   \end{center}
   \caption{Spatio-temporal intensity profiles of the  Bragg diffracted, reflected  beam (same as in Fig.  \ref{fig:E-wx}). Left: simulated with the FFT BPM method. Right: simulated intensity profile using the theory presented in \cite{PhysRevSTABYuri}. The dashed line has a slope -$c\cot{\theta_B}$ predicted theoretically in \cite{PhysRevSTABYuri}. Color scale units are arbitrary.  }
 \label{fig:E-tx}
   \end{figure}

Let us now consider a time-dependent radiation field: 
 \begin{equation}
   \psi^i_{txy}(t, x, y) = \frac{1}{\sqrt{2 \pi} \sigma_t} \psi^i_{xy} e^{(t - t_0)^2/2 \sigma_t^2}.
   \label{ref:tdp}
 \end{equation}
 The field in Eq. \eqref{ref:tdp} is Fourier transformed as $\psi^i_{\omega xy}(\omega, x, y) = \sum_{i=0}^N\psi^i_{t_ixy}(t_i, x, y) e^{i \omega t_i} \Delta t$, where $N$ is the number of grid points, thereby transforming a time-dependent diffraction problem into $N$ stationary tasks with a given $\omega_i$ and the spatial intensity distribution $\psi_{xy}$. After the simulation, the diffracted field is obtained via an inverse Fourier transform $\psi^h_{txy}(t, x, y) = \sum_{i=0}^N\psi^h_{\omega_ixy}(\omega_i, x, y) e^{-i \omega_i t} \Delta \omega$.
 Figs. \ref{fig:E-wx} and \ref{fig:E-tx} illustrate spatio-spectral and spatio-temporal responses of the $d=50$ $\mu$m thick $C^*(333)$ crystal reflection to the incident Gaussian beam of $\omega_0=12.8$ keV. Figure \ref{fig:E-tx} shows a perfect agreement between the theoretical calculations, presented in \cite{PhysRevSTABYuri}, and the results obtained by the FFT BPM method. In Fig. \ref{fig:E-tx} we plot the intensity to the power of 0.3 to enhance the contrast necessary to visualise x-ray wake fields created on a larger time scale.

  \subsection{Time-dependent problems: Deformed crystals }
 
  We will illustrate the simulation of x-ray reflection and transmission in a deformed crystal with an example related to the XFEL hard x-ray self-seeding scheme \cite{Amann2012, nam2021high}. In the self-seeding scheme, the crystal is placed between two undulator sections separated by a magnetic chicane. The first section produces an x-ray pulse by the self-amplified spontaneous emission (SASE) process. The SASE pulse consists of short temporal spikes. The average width of the spikes corresponds to the average spectrum bandwidth in the order of a few times $10^{-3}$. This is larger than the typical Darwin width. The narrow bandwidth seed pulse is created as a wakefield when the incident pulse propagates through the crystal. Here, for simplicity, we consider a 3D Gaussian pulse with the duration of 1 fs, the RMS transverse size of 30 $\mu$m, and the central photon energy of 9.831 keV. This pulse is impinging on a $d=120$ $\mu$m thick diamond crystal and experiencing the $C^*$(400) Bragg reflection for the incident angle of 45$^\circ$. The pulse energy of 50 $\mu$J creates an instantaneous temperature increase of 10 K in the center of the beam at the surface of the crystal.
  
  It has been demonstrated that Linac Coherent Light Source (LCLS) can produce pulse pairs separated by multiples of the RF period (0.35 ns), up to several hundreds of nanoseconds \cite{Decker2022}. The question we would like to address here is the influence of the first pulse on the properties of the wakefields of the second pulse.
  \begin{figure}
   \begin{center}
   \includegraphics[width=0.46\linewidth]{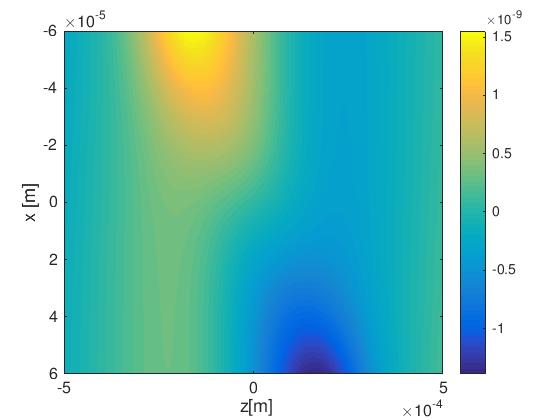}
      \includegraphics[width=0.46\linewidth]{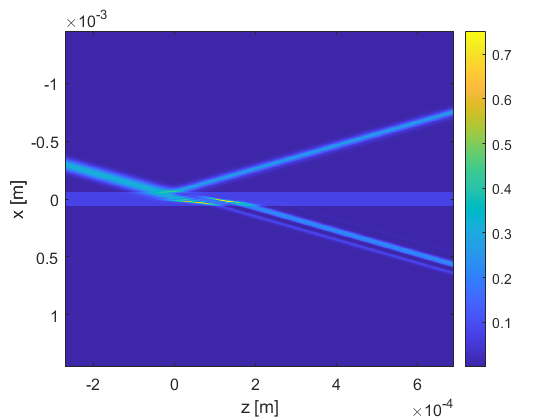}
   \end{center}
 
   \caption{Left: 2D visualization of deformation of the crystal caused by an intense XFEL pulse. The color scale, in meters, marks the displacement component perpendicular to the surface. Right: 2D visualization of Bragg reflection from the deformed crystal at the central frequency of the XFEL pulse.}
    \label{fig:stress-fenics}
   \end{figure}
  The deformation caused by the first pulse is simulated using an open-source finite element analysis (FEA) package FENICS \cite{Alns2015TheFP}. The main focus of this work is the demonstration of the application of the FFT BPM for describing the x-ray scattering process from a certain deformation of the crystal. Therefore, for simplicity, we have neglected the inertial vibrations of the crystal and used the steady-state solution, which is caused by the temperature profile of the crystal. In this example, we assume that the pulses are separated by 1 ns. The simulated displacement component, perpendicular to the crystal surface, is presented in Fig. \ref{fig:stress-fenics}. Here we have interpolated FEA results to match the grid used in the FFT BPM calculations (22 $\times$ 4267 $\times$ 1842). We are showing the section in the $xz$ plane at $y=0$, the symmetry plane of the scattering process. The results from the FFT BPM simulations are presented in Figs. \ref{fig:E-wx-deformed}, \ref{fig:E-tx-deformed-3D} and \ref{fig:E-tx-deformed-y0}. The left parts of the figures represent the spatio-frequency and spatio-temporal intensity distributions of the pulse scattered by the perfect crystal. The right part of the figures shows the effect of the deformation on the scattered pulse intensity. We have plotted intensities to the power of 0.6 in Fig \ref{fig:E-wx-deformed}, to the power of 0.2 in Fig. \ref{fig:E-tx-deformed-3D} and to the power of 0.3 in Fig. \ref{fig:E-tx-deformed-y0} to enhance the contrast ncessary to visualise weaker fields generated on a longer time scale. The surface plotted in Fig. \ref{fig:E-tx-deformed-3D} obeys the equation ${I_n}^{0.2}=0.33$ where ${I_n}$ is normalized intensity.

  When the pulse separation increases, the temperature profile of the crystal at the arrival of the second pulse relaxes. For example, for the separation of 200 ns, FENICS simulations show that the maximum temperature falls to about 0.3 degrees while the profile spreads out. Our simulations predict that at 200 ns separation, the wakefields' profile deformation is negligible. The only visible effect is the relative frequency shift of $3.5 \cdot {10}^{-7}$.
    In reality, the sudden energy deposition produces localized, trapped oscillations of the diamond crystal around a steady-state solution with a period of a few tens of nanoseconds and decay time in the microsecond range \cite{JWU}. These oscillations can influence the self-seeding of the XFEL operating at a high frequency where the separation of the consecutive pulses is comparable with the decay time of the oscillations. The study of these phenomena using the FFT BPM method will be published elsewhere.

 \begin{figure}
   \begin{center}
   \includegraphics[width=0.46\linewidth]{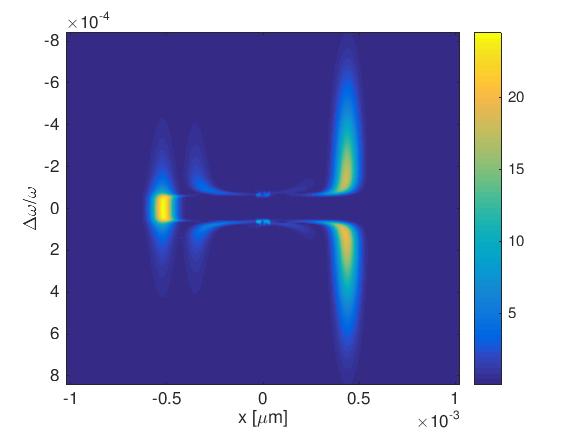}
      \includegraphics[width=0.45\linewidth]{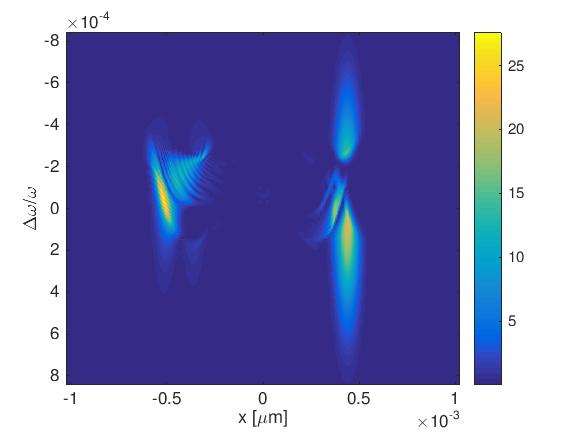}
   \end{center}
   \caption{ Spatio-spectral intensity profiles of the Bragg diffracted beam at $y=0$. Left: simulated with the FFT BPM method for the ideal crystal. Right: simulated with the presence of the deformation presented in Fig. \ref{fig:stress-fenics}. Color scale units are arbitrary.}
   \label{fig:E-wx-deformed}
   \end{figure}

 \begin{figure}
   \begin{center}
    \includegraphics[width=0.46\linewidth]{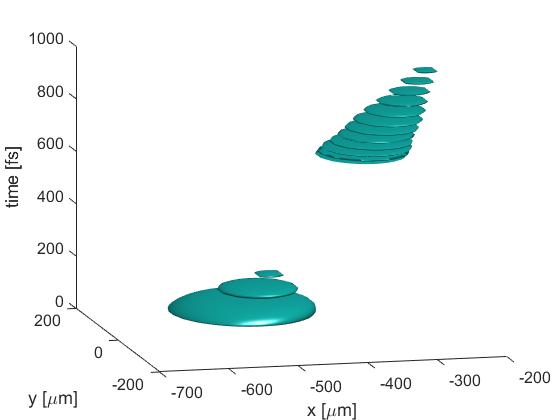}
    \includegraphics[width=0.46\linewidth]{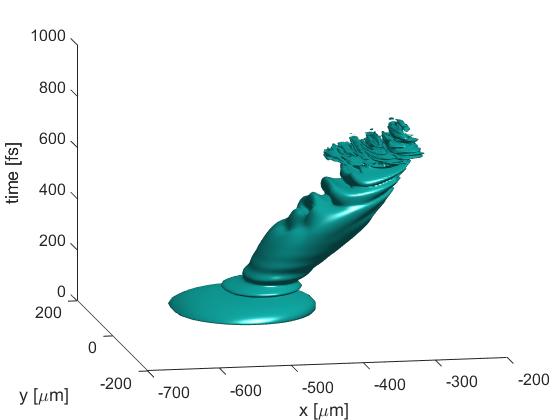}
   \includegraphics[width=0.46\linewidth]{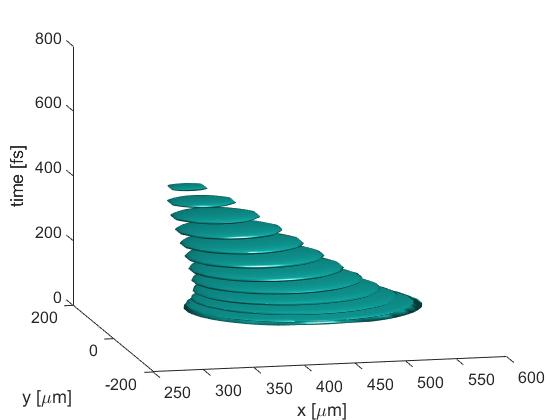}
         \includegraphics[width=0.46\linewidth]{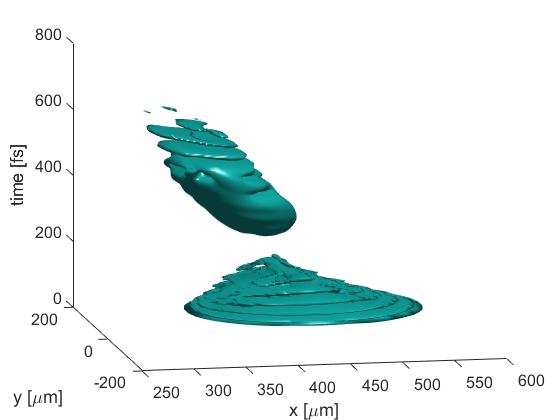}
   \end{center}
   \caption{Spatio-temporal intensity profiles of reflected (top) and transmitted (bottom) beam. Left: for the ideal crystal. Right: for  the deformed crystal shown in Fig. \ref{fig:stress-fenics}.} \label{fig:E-tx-deformed-3D}
   \end{figure}

\begin{figure}
   \begin{center}
   \includegraphics[width=0.46\linewidth]{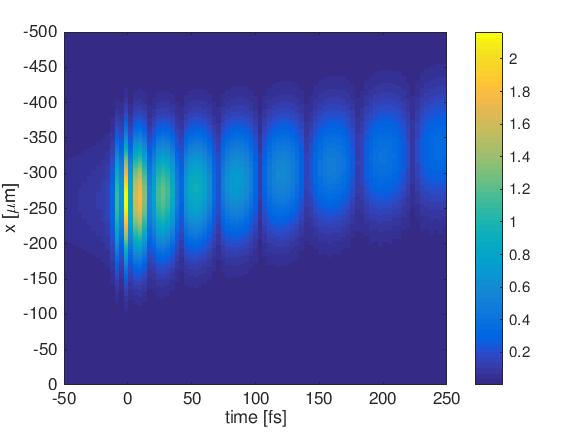}
      \includegraphics[width=0.46\linewidth]{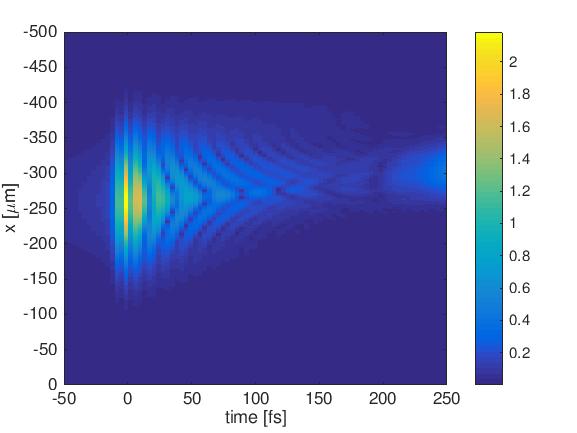}
    \end{center}
    \caption{Spatio-temporal intensity profiles of the Bragg diffracted and  transmitted beam at $y=0$. Left: for the ideal crystal. Right: for  the deformed crystal shown in Fig. \ref{fig:stress-fenics}. Color scale units are arbitrary.}
    \label{fig:E-tx-deformed-y0}
\end{figure}
   
\section{Summary}
 We demonstrated that our proposed FFT BPM method is accurate and numerically efficient in solving dynamical diffraction problems in complex optical systems. According to our knowledge this is the most numerically efficient method that allows to simulate in practice time dependent x-ray scattering from distorted crystals in 3D geometry. We have numerically shown that thermal deformations in high repetition rate XFEL self-seeding crystals result in spatio-spectral and spatio-temporal intensity profile distortions. Our method allows for accurate study of this phenomenon and can help design future self-seeding systems. Finally, we point out that application of FFT BPM is not limited to XFEL related optics, and can be extended to complicated monochromators, split-and-delay lines, and other hard x-ray instruments. Python programs used for simulation of the examples described  in this paper are available for download at \cite{fftbpm}.

\ack{Acknowledgements}

We thank James MacArthur, Zhengxian Qu (ex-SLAC) and Juhao Wu (SLAC) for insightful discussions and help with numerical simulations. 
This work is supported by the U.S. Department of Energy Office of Science under Contract No. DE-AC02-76SF00515. This research used resources of the National Energy Research Scientific Computing Center (NERSC), a U.S. Department of Energy Office of Science User Facility located at Lawrence Berkeley National Laboratory, operated under Contract No. DE-AC02-05CH11231 using NERSC award ERCAP0020725.

 

\bibliography{iucr}
\bibliographystyle{iucr}

\end{document}